\documentclass[12pt]{iopart}
\usepackage{iopams}  
\usepackage{amsfonts}
\usepackage{amssymb}
\usepackage{graphicx}
\usepackage{color}

\begin{document}
\title[Intrinsic and extrinsic measure for Brownian motion]{Intrinsic and extrinsic measure for Brownian motion}
\author{Pavel Castro-Villarreal}
\address{Centro de Estudios en F\'isica y Matem\'aticas B\'asicas y Aplicadas, Universidad Aut\'onoma de Chiapas, Carretera Emiliano Zapata, Km. 8, Rancho San Francisco, C. P. 29050, Tuxtla Guti\'errez, Chiapas, M\'exico}
\ead{pcastrov@unach.mx}
\begin{abstract}  
Based upon the Smoluchowski equation on curved manifolds three physical observables are considered for the Brownian displacement, namely,  geodesic displacement, $s$,  Euclidean displacement, $\delta{\bf R}$, and projected displacement $\delta{\bf R}_{\perp}$. The Weingarten-Gauss equations are used to calculate the mean-square Euclidean displacements in the short-time regime.  Our findings show that from an extrinsic point of view the geometry of the space affects the Brownian motion in such a way that the particle's diffusion is decelerated,  contrasting with the intrinsic point of view where dynamics is controlled by the sign of the Gaussian curvature [J. Stat. Mech. P08006 (2010)].  Furthermore, it is possible to give exact formulae for $\left<\delta{\bf R}\right>$ and $\left<\delta{\bf R}^{2}\right>$ on spheres and minimal surfaces, which are valid for all values of time. In the latter case, surprisingly, Brownian motion corresponds to the usual diffusion in flat geometries, albeit minimal surfaces have non-zero Gaussian curvature. Finally, the two-dimensional case is emphasized due to its close relation to surface self-diffusion in fluid membranes.
\end{abstract}

\vspace{2pc}
\maketitle

\section{Introduction}

Brownian motion occurs as a representation of several phenomena arising in various contexts, ranging from elementary particle physics \cite{Quark},  general relativity \cite{smerlak} and condensed matter \cite{review}. In the last decade, there has been a lot of interest in the study of diffusive processes on ma\-ni\-folds motivated by problems coming  from biophysics \cite{Adler, Domanov}. For instance, the transport processes occurring on a bio\-lo\-gi\-cal cell are interesting and complex phenomena. In particular, the motion of an integral protein through the plasma membrane  has been approached from complementary points of view. The most basic of these is grounded on the Smoluchowski equation for a point particle on the membrane \cite{Aizenbud}. Further approaches consider the membrane's  thermal fluc\-tua\-tions \cite{Gustaffson, Seifert}, as well as dynamical fluctuations coupled to the stochastic motion of the protein \cite{Naji}, and finite-size effects of the protein \cite{Naji2}. It is noteworthy to mention that such membrane is considered as a two-dimensional curved surface because on mesoscopic scales most of its  mechanical properties are captured by geometrical degrees of freedom, with the membrane bending energy being quadratic in the extrinsic curvature \cite{helfrich}. 

In this work, we use Smoluchowski's equation to study the geometric component of the diffusion process over a curved surface.  In ge\-ne\-ral, we focus on the following questions: what observables are appropriate to measure the displacement of a Brownian particle? and how does the curvature affect the mean-value of these observables?  By analogy with Euclidean spaces it is not difficult to realize that a free particle in curved space moves from one point to another through a geodesic distance, of length $s$, \cite{Faraudo}.  However, it would be difficult to measure it in an experiment because, usually, the data extracted correspond to the diffusion on a projected plane. Thus, it is convenient to introduce Euclidean and projected displacements as alternative observables to the geodesic displacement. The Euclidean displacement $\delta{\bf R}:={\bf X}-{\bf X}_{0}$ is the vector difference between two points on the surface, and the projected Euclidean displacement $\delta{\bf R}_{\perp}:=\pi(\delta{\bf R})$,  is the projection of the Euclidean displacement onto the $xy$ plane, where $\pi$ is the projection map. These three quantities, $s$, $\delta{\bf R}$ and $\delta{\bf R}_{\perp}$ reduce to the same one for flat geometries.  
 
On one hand, the Euclidean displacement has been used in \cite{Holyst} to analyze the Brownian dynamics in various  geometries such as spheres, cylinders, periodic nodal surfaces and the so called P, D, G minimal surfaces, where, it is noteworthy to mention, that the structure of lyotropic surfactant phases like the $L_{3}$ (sponge) phase is well understood from an experimental point of view \cite{Amir} using a periodic minimal surface model \cite{Englobm}. On the other hand, the projected displacement has been proposed in \cite{Gustaffson} and  \cite{Seifert} to be the proper quantity to compare experimental data with for the lateral diffusion on fluid vesicles.  In addition, $\delta{\bf R}$ has also been implemented as the end-to-end distance to study how polymers wrap into a curved interface \cite{Muthukumar}. Additionally, the Euclidean displacement  turns out to be the same quantity as the rotational displacement  of rods  \cite{Dhont} due to the analogy between the  diffusion of a single long rod and the diffusion of a single particle  on the unit sphere. 

As we have mentioned in \cite{Castro-2010}, from a theoretical viewpoint, the Brownian motion can be used to probe the geo\-me\-try of the manifold in the same spirit of Kac's famous question: {\it can one hear the shape of a drum?} \cite{Kac}. Consequently, the mean-square geodesic displacement  will probe the intrinsic properties of the surface whereas the mean-square Euclidean displacements will inherit the extrinsic properties of the surface. Here, we use the same technique used in \cite{Castro-2010} to compute the curvature effects on $\left<\delta{\bf R}\right>$, $\left<\delta{\bf R}^{2}\right>$ and $\left<\delta{\bf R}^{2}_{\perp}\right>$, but now with extensive use of the Weingarten-Gauss equations. In particular, we provide examples to study the diffusion on the sphere, catenoid and Clifford torus from  both, intrinsic and extrinsic, points of view.  Furthermore, it is observed that for certain conditions,  the expectation values, $\left<\delta{\bf R}\right>$ and $\left<\delta{\bf R}^{2}\right>$, can be computed exactly for the sphere and the minimal surfaces for all values of time. It is shown that the diffusion on minimal surfaces measured with the observable $\delta{\bf R}$ corresponds to the free usual diffusion in flat geometries in agreement with the cubic minimal surfaces already studied in \cite{Holyst} and \cite{Anderson}. 

The paper is organized as follows:  In section 2, we introduce the Smoluchowski equation and the displacement observables on curved manifolds. In section 3, we present the operator method used to evaluate the expectation values of observables. In particular, we review  the short-time regime of $s^{2}$, and we study the short-time regime of $\delta{\bf R}^2$, $\delta{\bf R}$ and $\delta{\bf R}^2_{\perp}$.  In section 4, we find exact results for $\left<\delta{\bf R}^2\right>$ and $\left<\delta{\bf R}\right>$ in the cases of a sphere and minimal surfaces. In addition, in section 5, we provide illustrative examples of diffusion on sphere, catenoid and Clifford torus. Finally, in section 6, we summarize our main results and give our concluding perspectives.

\section{Smoluchowski equation and observables on curved manifolds}
\label{sec:2}
In this section, we introduce the simplest model to study Brownian motion on curved manifolds.  This is a direct generalization of the Smoluchowski equation  on Euclidean spaces \cite{Dhont}, which basically consists of replacing the Euclidean Laplacian by the Laplace-Beltrami operator $\Delta_{g}$. This operator is often used to describe how a substance diffuses over a curved manifold. Also we may think of it as heat diffusing on manifolds or a polymer embedded on  curved interfaces \cite{Muthukumar}. Furthermore, it can be used to determine the quantum propagator of a free particle on curved spaces \cite{Chaichian}. 

For a single particle diffusion over a manifold $\mathbb{M}$\footnote{See appendix A for relevant notation.} with dimension $d$, we are interested in the probability density  $P: \mathbb{M}\times\mathbb{M}\times\mathbb{R}^{+}\to\mathbb{R}^{+}$ such that $P\left(x,x^{\prime},t\right)dv$  is the probability to find a diffusing  particle in the volume element $dv$  centered in $x$, at time $t$, when the particle started in $x^{\prime}$ at $t=0$ \footnote{ $P$ also can be interpreted as the density probability  for $n$ non-interacting particles diffusing on an manifold ${\rm N}$ provided that $\mathbb{M}=\otimes^{n}_{i=1}{\rm N}$.}. This probability distribution $P$  is governed by the Smoluchowski equation 
\begin{eqnarray}
\frac{\partial P\left(x,x^{\prime},t\right)}{\partial t}=D\Delta_{g}P\left(x,x^{\prime},t\right),
\label{diff.Eq}
\end{eqnarray}
where $D$ is the diffusion coefficient, and satisfies the initial condition at time $t\to0$, 
\begin{eqnarray}
 \lim_{t\to 0}P\left(x,x^{\prime},t\right)=\delta^{d}\left(x-x^{\prime}\right)/\sqrt{g}.
\label{ini.cond}
\end{eqnarray}
For most geometries an exact solution for the probability distribution is not known. However, for short times there is a formal power series solution for $P\left(x,x^{\prime},t\right)$ in terms of the Minakshisundaram-Pleijel coefficients \cite{Pleijel}, which depends on both $x$ and $x^{\prime}$ \cite{Denjoe}.

In order to get some insight about  Brownian motion we often look at mean-values of observables ($\mathcal{O}\left(x\right)$)
defined on the manifold such that 
\begin{eqnarray}
\left<\mathcal{O}\left(x\right)\right>=\int_{\mathbb{M}} dv~ \mathcal{O}\left(x\right)P\left(x,x^{\prime}, t\right)
\end{eqnarray}
is well-defined for all points $x^{\prime}$ in the manifold and for all time values; in particular, for $d=2$, $dv\equiv dA$ is the area element of the surface and $\left<\mathcal{O}\left(x\right)\right>$ is the mean-value of an observable on the surface. Note that $\left<\mathcal{O}\left(x\right)\right>$ depends  on the initial position ${ x}^{\prime}$.  We will restrict ourselves to the case of functions that are  $C^{\infty}\left({\mathbb{M}}\right)$. In general, the structure of  $\left<\mathcal{O}\left(x\right)\right>$ depends on the symmetries of the space and the boundary conditions imposed on $P$. For instance, for a free Brownian particle suspended in the bulk of an homogenous fluid the mean-displacement, $\left<\delta{\bf R}\right>=0$, and the mean-square displacement, $\left<\delta{\bf R}^{2}\right>=2dDt$, are rotational and traslational invariants. Now, we use the geodesic or Riemann distance, $s$, to measure how a Brownian particle moves on a curved manifold. The expectation value of $s^2$ measures how the intrinsic geometry influences Brownian motion. Alternatively, the displacement of the particle can also be measured using the Euclidean $\delta{\bf R}$  and projected displacements, $\delta{\bf R}_{\perp}$. From this point of view $\delta{\bf R}$ is defined by $\delta{\bf R}:\mathbb{M}\to {\mathbb R}^{d+1}$ given by  $\delta{\bf R}={\bf X}-{\bf X}_{0}$ for  all points ${\bf X}\left(x_{1}, \cdots,x_{d}\right)\in \mathbb{M}$, where ${\bf X}$ is  a parametrization of the manifold.  In addition, we define $\delta{\bf R}_{\perp}$ by the projection map $\delta{\bf R}_{\perp}\equiv\pi\left(\delta{\bf R}\right)$\footnote{ $\pi:\mathbb{R}^{d+1}\to\mathbb{R}^{d}$ is defined as usual by $\pi\left({\bf V}\right)={\bf v}$ for ${\bf V}=\left({\bf v}, v_{0}\right)\in\mathbb{R}^{d+1}$}. In particular, for a domain of $\mathbb{M}$ covered with one coordinate neighborhood we use the Monge parametrization ${\bf X}=\left({\bf x},h\left({\bf x}\right)\right)$, where  $h\left({\bf x}\right)$ is the height function for ${\bf x}\in \mathbb{R}^{d}$. Thus, in this representation we have $\delta{\bf R}_{\perp}={\bf x}-{\bf x}_{0}$.

\section{Expectation values of observables}
\label{sec:3}

The general problem is to find the mean values for  $s^2$, $\delta{\bf R}$, $\left|\delta{\bf R}\right|^{2}$ and $\left|\delta{\bf R}_{\perp}\right|^{2}$ for any embedded manifold.  In principle, these expectation values can be eva\-lua\-ted through the formal series solution of the diffusion equation in terms of the Minakshisundaram-Pleijel coefficients mentioned above.  However, here we use an operator method, introduced in \cite{Castro-2010},  inspired by the original calculations made by Perrin in his seminal paper about Brownian motion on spheres \cite{Perrin}. The idea consists in approximating the expectation values by a Taylor polynomial in the time variable.  In addition, we consider that $P\left(x, x^{\prime},t\right)$ and $\nabla^{a}P$ vanish at the boundary of the manifold. Using these conditions we are allowed to write
\begin{eqnarray}
\left.\frac{\partial^{k}\left<\mathcal{O}\left(x\right)\right>}{\partial t^{k}}\right|_{t=0}=\left.D^{k}\Delta^{k}_{g}\mathcal{O}\left(x\right)\right|_{x=x^{\prime}}, 
\end{eqnarray} 
valid for differentiable observables $\mathcal{O}\left(x\right)$ (see Appendix B). In addition, assuming that  $\left.\partial^{k}\left<\mathcal{O}(x)\right>/\partial t^{k}\right|_{t=0}$ are well defined on $\mathbb{R}^{+}$, for a given  $\mathbb{M}$, we define the remainder $\epsilon_{n}\left(t\right)$ by
\begin{eqnarray}
\left<\mathcal{O}\left(x\right)\right>=\sum^{n}_{k=0}\frac{G^{{\mathcal{O}}}_{k}}{k!}\left(Dt\right)^{k}+\epsilon_{n}\left(t\right),
\label{formula}
\end{eqnarray}
where the terms $G^{{\mathcal{O}}}_{k}\equiv\left.\Delta^{k}_{g}\mathcal{O}\right|_{x=x^{\prime}}$ are purely geometric factors. Now for a given observable  the difficulty lies in evaluating the terms $G^{{\mathcal{O}}}_{k}$.  The equation (\ref{formula}) is very useful to access the short-time regime of the Brownian motion for the general manifold case, but  it can also be used to find closed formulae valid for all values of time for some specific manifolds and observables.

In what follows, we give a third order polynomial approximation for the mean values of the observables mentioned above. To this order, the Taylor polynomial is an approximation of $\left<\mathcal{O}\left(x\right)\right>$ with an error of $\delta_{4}=\epsilon_{4}/C_{4}$ if the time $t$ satisfies $t<\tau_{G}\equiv\left(4!\delta_{4}\right)^{1/4}$, where $C_{4}$ is a constant defined at Appendix D.  This later inequality also defines the short-time regime.

\subsection{Intrinsic Brownian motion at short-time regime }
 
The mean-value of $s^2$ captures intrinsic geometrical data of the manifold and says how the geo\-me\-try causes a change in the standard diffusive behavior. The geometric factors for this observable cannot be written, in general, in a closed form for all values of $k$. However, in \cite{Castro-2010} we have computed these factors for the first three values, $k=1,2,3$. The result is a third order Taylor polynomial, 
\begin{eqnarray}
\left<s^{2}\right>=2dDt-\frac{2}{3}R_{g}\left(Dt\right)^{2}&+&\frac{1}{3!}\left[\frac{8}{15}R^{ab}R_{ab}\right.-\left.\frac{16}{45}R^{abcd}\left(R_{dbca}+R_{dcba}\right)\right.\nonumber\\
&-&\left.\frac{16}{5}\left(\nabla^{a}\nabla^{b}+\frac{1}{2}g^{ab}\Delta_{g}\right)R_{ab}\right]\left(Dt\right)^{3}+\cdots\nonumber,\\
\label{mean-square}
\end{eqnarray}
with coefficients taking values at the points of the manifold. This result shows how the  mean-square GD deviates from the planar expression by terms which are invariant under general coordinate transformations. All these terms are built with $O(d)$ invariant combinations of the Riemann tensor. In particular,  for regular surfaces  in $\mathbb{R}^{3}$, the Riemann tensor components are $R_{abcd}=\frac{K_{G}}{4}\left(g_{ac}g_{bd}-g_{ad}g_{bc}\right)$, where $K_{G}$ is the Gaussian curvature. Note that for developable surfaces, $K_{G}=0$, the mean-square geodesic displacement behaves like the usual diffusion in the short-time regime \cite{Faraudo}. 

\subsection{Extrinsic Brownian motion at short-time}

\noindent For the Brownian motion over manifolds of dimension $d$ (embedded in $\mathbb{R}^{d+1}$) we are interested in the expectation values of $\mathcal{O}_{2}\equiv\left|\delta{\bf R}\right|^{2}$ and $\delta {\bf R}$. In what follows let us recall that $K_{ab}$ are the components of the second fundamental form and $K=g^{ab}K_{ab}$ is the mean curvature of the hypersurface.  

In this case, also,  the geometric factors   cannot be evaluated in a closed form for all integers $k$; we will calculate them only for $k=1,2,3$.  Clearly, for $k=1$ we have $G^{\mathcal{O}_{2}}_{1}=\Delta_{g}\left|\delta{\bf R}\right|^{2}=2\nabla_{a}\left(\delta{\bf R}\cdot{\bf e}^{a}\right)$. Using the trace of metric tensor is $g_{~a}^{a}=d$, and the Weingarten-Gauss equation, (\ref{Weingarten-Gauss}), we get $G^{\mathcal{O}_{2}}_{1}=\left.\left(2d-2K\delta{\bf R}\cdot{\bf N}\right)\right|_{\delta{\bf R}=0}=2d$.
In a similar way, by straightforward calculation, we obtain $G^{\mathcal{O}_{2}}$ and $G^{\mathcal{O}_{3}}$ (see appendix C). Hence the mean-square Euclidean displacement $\left<\delta{\bf R}^{2}\right>$ is given by 
\begin{eqnarray}
\left<\delta{\bf R}^{2}\right>= 2dDt&-&K^{2}\left(Dt\right)^{2}\nonumber\\&-&\frac{1}{3}\left[K\Psi\left(K\right)+\Delta_{g}\left(K^{2}\right)+2\nabla_{b}J^{b}\left(K\right)\right]\left(Dt\right)^{3}+\cdots. 
\label{resultado1}
\end{eqnarray}
In addition, following the same procedure used just above we find
the expectation value of $\delta{\bf R}$,
{\small\begin{eqnarray}
\left<\delta{\bf R}\right>=-K{\bf N}Dt&-&\frac{1}{2}\left(\Psi\left(K\right){\bf N}+J^{b}\left(K\right){\bf e}_{b}\right)\left(Dt\right)^{2}-\frac{1}{6}\left\{\left[\Delta_{g}\Psi\left(K\right)\right.\right.\nonumber\\&-&\left.\left.\Psi\left(K\right) K_{cd}K^{cd}-\left(2K_{cd}\nabla^{c}J^{d}\left(K\right)+J^{b}\left(K\right)\nabla_{a}K^{a}_{~b}\right)\right]{\bf N}\right.\nonumber\\&+&\left.\left[-2\nabla^{c}\Psi\left(K\right) K_{c}^{~b}-\Psi\left(K\right)\nabla_{a}K^{ab}+J^{b}\left(K\right)K_{cb}K^{cd}\right.\right.\nonumber\\&-&\left.\left.\Delta_{g}J^{b}\left(K\right)\right]{\bf e}_{b}\right\}\left(Dt\right)^{3}+\cdots,
\label{resultado2}
\end{eqnarray}}
where the scalar $\Psi\left(K\right)$ and vector $J^{a}\left(K\right)$ are defined as 
\begin{eqnarray}
\Psi\left(K\right)=\Delta_{g}K-KK_{ab}K^{ab},~~~~~~~~
J^{a}\left(K\right)=K\nabla_{b}K^{ba}+2K^{ab}\nabla_{b}K.
\end{eqnarray}
These results, (\ref{resultado1}) and (\ref{resultado2}), show how the Euclidean mean-values  deviate from its flat counterpart. This deviation is also given by terms invariant under general coordinate transformations,  but now they are built in terms of the $O(d)$-invariant of  the extrinsic curvature tensor $K_{ab}$, i.e., they are referred to the ambient space where the hypersurface is embedded. 

As in the geodesic displacement, in a local neighborhood, the mean-square Euclidean displacement reproduces the standard Einstein kinematical relation. Additionally, it can be noted that for non-zero short-times, $t\ll\tau_{G}$, we still have a contribution coming form the curvature: $\left<\delta{\bf R}\right>\approx-K{\bf N}Dt$. At these short-times, the normal contribution came about because in average tangent directions cancel out each other by symmetry. Nevertheless, as soon as the particle reaches the boundary of the local neighborhood, the tangent components start to contribute.

\subsection{ Projected Brownian motion at short-times }

In a subspace $\mathbb{R}^{d}$ ($\subset\mathbb{R}^{d+1}$),  the projected displacement is defined by $\delta{\bf R}_{\perp}\equiv\pi\left({\bf X}\right)={\bf x}$, for ${\bf x}_{0}=0$.  In order to compute the mean-square PD,  $\left<\delta{\bf R}^{2}_{\perp}\right>$, let us write $\mathcal{O}_{4}\equiv {\delta\bf R}^{2}_{\perp}={\bf X}^2-h^{2}$, therefore the geometric factors satisfy 
\begin{eqnarray}
G^{\mathcal{O}_{4}}_{k}=G^{\mathcal{O}_{2}}_{k}-\left.\Delta^{k}_{g}h^{2}\right|_{{\bf X}=0}.
\end{eqnarray}
Thus, for the geometric factors with $k=1,2,3$ we have to calculate $\Delta_{g}h^{2}$, $\Delta^{2}_{g}h^{2}$ and $\Delta^{3}_{g}h^{2}$. For $k=1$, we have $\Delta_{g}h^{2}=2\nabla_{a}h\nabla^{a}h$. Using the expression for the metric in this parametrization the latter factor can be written as $\Delta_{g}h^{2}=2\left(\partial h\right)^{2}/\left(1+\left(\partial h\right)^{2}\right)$ and using the normal vector ${\bf N}$ in this parametrization\footnote{See appendix A.1.3.} we find
\begin{eqnarray}
\Delta_{g}h^{2}=2\left(1-N^{2}_{z}\right).
\end{eqnarray}
The terms $\Delta^{2}_{g}h^{2}$ and $\Delta^{3}_{g}h^{3}$ are left expressed in covariant form. By straightforward calculation we find
\begin{eqnarray}
\Delta^{2}_{g}h^{2}&=&4\left(\nabla^{a}\nabla^{b}h\right)\left(\nabla_{a}\nabla_{b}h\right)+4\left(\Delta_{g}h\right)^{2}+12\left(\Delta_{g}\nabla_{a}h\right)\left(\nabla^{a}h\right),\nonumber\\
\nonumber\\
\Delta^{3}_{g}h^{2}&=&8\left(\nabla^{a}\nabla^{b}\nabla^{c}h\right)\left(\nabla_{a}\nabla_{b}\nabla_{c}h\right)+32\left(\nabla^{a}\nabla^{b}\Delta_{g}h\right)\left(\nabla_{a}\nabla_{b}h\right)\nonumber\\&+&20\left(\nabla_{a}\Delta^{2}_{g}h\right)\left(\nabla^{a}h\right)
+12\left(\nabla_{a}\Delta^2_{g}h\right)\nabla^{a}h+8\left(\Delta^{2}_{g}h\right)\left(\Delta_{g}h\right).
\label{ora}
\end{eqnarray}
Note that $\nabla_{a}$ is the covariant derivative compatible with the metric. As in the previous cases, the mean-square projected displacement is written as 
$\left<\delta{\bf R}^{2}_{\perp}\right>\approx 2dD_{proj}t+\frac{1}{2!}G^{\mathcal{O}_{4}}_{2}\left(Dt\right)^{2}+\frac{1}{3!}G^{\mathcal{O}_{4}}_{3}\left(Dt\right)^{3}+\cdots
$. For very short-times $t\ll \tau_{G}$, the mean-square projected displacement has the usual diffusion behavior in flat geometries, but with a new diffusion coefficient 
\begin{eqnarray}
D_{proj}=\frac{D}{d}\left(d-1+N^{2}_{z}\right).
\label{Diff}
\end{eqnarray}
 This means that for the Brownian motion observed from the projected sub-space the diffusion is reduced, since $D_{proj}$ is smaller than $D$. This modification is just a geometrical effect due to the point of view from where  the Brownian motion is measured. 
 
\vskip0.5em\noindent{\it Remark}. The result (\ref{Diff}) has been obtained by different methods for $d=2$ in  \cite{Gustaffson} and \cite{Seifert} within the context of lateral diffusion of integral proteins in biomembranes.  In these works they also consider  membrane ther\-mal fluctuations. For instance in \cite{Seifert}, within the Helfrich-Canham model \cite{helfrich} for fluid membranes the effective value of the diffusion coefficient is computed, (\ref{Diff}), when thermal fluctuations are considered. In this case, the coefficients $G^{\mathcal{O}_{4}}_{k}$ will be affected by thermal fluctuations; this will be taken into consideration elsewhere.

\section{Special cases of Brownian motion on spheres, cylinders and minimal surfaces} 

We consider now the whole  series (\ref{formula}) for the set of observables $\left\{\delta{\bf R}, \delta{\bf R}^{2},\delta{\bf R}^{2}_{\perp}\right\}$. For these ob\-ser\-va\-bles, as we shall see,   we are able to give exact and closed formulae valid for all time values in the cases of spheres and minimal surfaces. Let us start recalling the following result. Let  $\mathcal{O}\left(x\right)$ be an eigenfunction of the Laplace-Beltrami operator  $\Delta_{g}$ with eigenvalue $-\lambda$, then the expectation value of $\mathcal{O}\left(x\right)$ is given by
\begin{eqnarray}
\left<\mathcal{O}\left(x\right)\right>=\mathcal{O}\left(x^{\prime}\right)\exp\left(-\lambda Dt\right).
\label{resA}
\end{eqnarray}
In addition, for those functions $\mathcal{O}\left(x\right)\in C^{\left(2\right)}\left(\mathbb{M}\right)$ such that  $ \Delta_{g}\mathcal{O}\left(x\right)=C\equiv{\rm constant}$ for each point on $\mathbb{M}$ we have  \begin{eqnarray}
\left<\mathcal{O}\left(x\right)\right>=\mathcal{O}\left(x^{\prime}\right)+CDt.
\label{resB}
\end{eqnarray}
The proof of these results is given in Appendix D. The spheres ($S^{d}$) admit a parametrization that satisfies the condition in the first of these results, (\ref{resA}), namely, that $-\Delta_{g}{\bf X}=\frac{d}{R^2}{\bf X}$. In fact, if a $d-$dimensional manifold (embedded in $\mathbb{R}^{d+1}$) has a parametrization ${\bf X}$ that is an eigenfunction of $\Delta_{g}$ then it has to be a piece of sphere or a piece of a minimal hypersurface (with eigenvalue $\lambda=0$).  These results are useful to prove a general structure for expectation values of $\delta{\bf R}$,  $\delta{\bf R}^{2}$ and $\delta{\bf R}^{2}_{\perp}$ for  minimal hypersurfaces  ($K=0$) and spheres $S^{d}$.  These properties  imply the following results
\vskip0.5em
\noindent {\it\bf Result  1}. For each minimal hypersurface of dimension $d$ the expectation values of $\delta{\bf R}$ and $\delta{\bf R}^{2}$, with respect to $P$,  are given by
\begin{eqnarray}
\label{ms1}
\left<\delta{\bf R}\right>&=&0,\\
\label{ms2}
 \left<\delta{\bf R}^{2}\right>&=&2dDt, 
\end{eqnarray} 
for all values of time $t$.\\
\vskip0.5em
\noindent{\it Proof.} Since the embedded manifold is a minimal hypersurface we have $\Delta_{g}{\bf X}=0$ and by result (\ref{resA}) we get $\left<{\bf X}\right>={\bf X}_{0}$ $\Leftrightarrow$ $\left<\delta{\bf R}\right>=0$. Now, for the mean-square ambient displacement, let ${\bf X}$ be a parametrization of the minimal hypersurface then by the Weingarten-Gauss equations $\Delta_{g}{\bf X}^{2}=2d$, therefore by result (\ref{resB}) we get $\left<{\bf X}^{2}\right>={\bf X}^{2}_{0}+2dDt$, that is $\left<\delta{\bf R}^{2}\right>=2dDt$. 
\vskip0.5em
\noindent {\it Remark.} These results, (\ref{ms1}) and (\ref{ms2}),   are consistent with the general formula at the short-time regime  (\ref{resultado1}) and (\ref{resultado2}), respectively. In addition, note that for $d=2$ this is in agreement with the particular P, D, G minimal surfaces explicitly studied in \cite{Holyst} and \cite{Anderson}. 
\vskip0.5em
\noindent Now, let us give a formula for the expectation value of $\delta{\bf R}^{2}_{\perp}$ in the case of  minimal surfaces. Thus we get the following 

\vskip0.5em
\noindent {\it\bf Result  2}. For each minimal hypersurface of dimension $d$ the expectation values of  $\delta{\bf R}^{2}_{\perp}$, with respect to $P$,  is given by
\begin{eqnarray}
 \left<\delta{\bf R}^{2}_{\perp}\right>=2D_{\rm proj}t-\sum^{\infty}_{k=2}\left(\nabla^{a_{1}}\cdots\nabla^{a_{k}}h\right)\left(\nabla_{a_{1}}\cdots\nabla_{a_{k}}h\right)\frac{\left(2Dt\right)^{k}}{k!}, 
 \label{projMS}
\end{eqnarray} 
for all values of time $t$, where $h$ is the height function of the Monge representation and $D_{\rm proj}=\frac{D}{2}\left(1+N^{2}_{z}\right)$.\\
\vskip0.5em
\noindent{\it Proof.} Using the Monge parametrization,  ${\bf X}\left({\bf x}\right)=\left({\bf x},h\left({\bf x}\right)\right)$, we have $\delta{\bf R}^{2}_{\perp}={\bf X}^2-h^2$. Now, because the surface is minimal then by {\bf Result 1}, 
$\left<\delta{\bf R}^{2}_{\perp}\right>=4Dt-\left<h^2\right>$. Next, we have to compute $\Delta^{k}_{g}h^2$, by inspection this is equivalent to $2^{k}\left(\nabla^{a_{1}}\cdots\nabla^{a_{k}}h\right)\left(\nabla_{a_{1}}\cdots\nabla_{a_{k}}h\right)$, the rest of the terms in $\Delta^{k}_{g}h^2$ involve $\Delta_{g}h$ which is zero for a minimal surface. 
\vskip0.5em
\noindent In a similar fashion we can compute the mean values of $\delta{\bf R}$ and $\delta{\bf R}^{2}$ for spheres ($S^{d}$) and compare with known results (for instance in \cite{Muthukumar}.) We get the following
\vskip0.5em
\noindent
\noindent {\it\bf  Result 3}. For spheres $S^{d}$ the expectation values of $\delta{\bf R}$ and $\delta{\bf R}^{2}$, with respect to $P$,  are given by
\begin{eqnarray}
\label{dRt}
\left<\delta{\bf R}\right>&=&{\bf X}_{0}\left(\exp\left(- \frac{d}{R^{2}}Dt\right)-1\right),\\
 \left<\delta{\bf R}^{2}\right>&=&2R^{2}\left(1-\exp\left(-\frac{d}{R^{2}}Dt\right)\right),
\label{dR2t}
\end{eqnarray} 
for all values of time.\\
\vskip0.5em 
\noindent {\it Proof}. Since the manifold is a sphere $S^{d}$ then there is a pa\-ra\-me\-tri\-za\-tion ${\bf X}$ such that $-\Delta_{g}{\bf X}=\frac{d}{R^2}{\bf X}$, that is ${\bf X}$ is an eigenfunction of $\Delta_{g}$ with eigenvalue $-\frac{d}{R^2}$, then by result (\ref{resA}) we get $\left<{\bf X}\right>={\bf X}_{0}\exp\left(-\frac{d}{R^2}Dt\right)$, where ${\bf X}_{0}$ is the starting point. Now, since the manifold is a sphere, $\Delta_{g}{\bf X}^{2}=0$, then $\left<\delta{\bf R}^{2}\right>=\left<{\bf X}^{2}\right>+\left<{\bf X}^{2}_{0}\right>-2\left<{\bf X}\right>\cdot{\bf X}_{0}$, hence substituting the result for $\left<{\bf X}\right>$ we get the claimed result.\\

\noindent Note that for sphere $S^{d}$ equations (\ref{dRt}) and (\ref{dR2t}) in the short-time regime reproduce (\ref{resultado1}) and (\ref{resultado2}), respectively. Result 3 was found in \cite{Muthukumar} and \cite{Holyst} by alternative methods. 
 Now, let us consider a $d$-dimensional infinite cylinder with radius $R$. This cylinder can be thought as ${\rm Cyl}\equiv S^{d-1}\times{\mathbb{R}}$. In this case, the embedding functions can be written in terms of that $(d-1)$-dimensional sphere  ${\bf X}=\left({\bf X}_{S^{d-1}}, z\right)$, where  $z\in\mathbb{R}$. The metric  can be written as $g_{ab}={\rm diag}\left(1, g_{ij}\right)$, where $g_{ij}$ is  the metric of $S^{d-1}$ and the Laplace-Beltrami operator in this case is given by $\Delta_{{\rm Cyl}}=\Delta_{S^{d-1}}+\partial^{2}/\partial z^{2}$. The starting point is chosen to be ${\bf X}_{0}=\left(1, 0, \cdots,0\right)$, thus we have that the expectation values of $\delta{\bf R}$ and $\delta{\bf R}^{2}$, with respect to $P$, for cylinders ${\rm Cyl}$ are given by
\begin{eqnarray}
\left<\delta{\bf R}\right>&=&{\bf X}_{0}\left(\exp\left(-\left(d-1\right)\frac{Dt}{R^{2}}\right)-1\right),\\
\left<\delta{\bf R}^{2}\right>&=&2Dt+2R^{2}\left(1-\exp\left(-\left(d-1\right)\frac{Dt}{R^{2}}\right)\right)
\end{eqnarray}  
for all values of time.

In a similar way, we can compute $\delta{\bf R}^{2}_{\perp}$ for a hemisphere of radius $R$. Thus,  let us take a hemisphere of radius $R$ and let  $\Pi$ be the projected subspace from this hemisphere onto $xy$-plane.  Thus a parametrization of this hemisphere is  ${\bf X}:\Pi\to\mathbb{R}^{3}$ defined by 
${\bf X}\left(\varphi, \rho\right)=\delta{\bf R}_{\perp}+\hat{\bf k}\sqrt{R^2-\rho^{2}} $, where $\delta{\bf R}_{\perp}=\left(\rho\cos\varphi, \rho\sin\varphi, 0\right)$. Then we get the following\\

\noindent {\it\bf  Result 4} For a hemisphere of radius $R$ the expectation values of $\delta{\bf R}^{2}_{\perp}$, with respect to $P$,  is given by
\begin{eqnarray}
\left<\delta{\bf R}^2_{\perp}\right>=\frac{2}{3}R^2\left(1-e^{-\frac{6Dt}{R^2}}\right),
\label{dR2perp}
\end{eqnarray}
for all values of time.\\
\vskip0.5em 
\noindent {\it Proof}.  In this case, we can verify that $\Delta_{g}\left(\delta{\bf R}^{2}_{\perp}\right)=4-\frac{6}{R^2}\rho^2$ for all values of $\rho\in\mathbb{R}^{+}$, thus it is not difficult to show through a straightforward calculation that $\left.\Delta^{k}_{g}\left(\delta{\bf R}^{2}_{\perp}\right)\right|_{\rho=0}=4\left(-\frac{6}{R^2}\right)^{k-1}$ for all naturals $k\neq 0$. Also, for this observable, $\delta{\bf R}^{2}_{\perp}$, we can verify that the remainder $\epsilon_{n}$ goes to zero for large $n$. \\

\noindent At short-times this expression reproduces the general structure for the projected mean-square displacement. In particular, the diffusion coefficient does not change since $N_{z}=1$ at $\rho=0$. 

\section{Illustrative examples}
In this section we provide examples to study  diffusion on the sphere, catenoid and Clifford torus. The geometrical dimensions of these surfaces are shown in figure (1).  Now, in order to quantify how much the mean-square displacements deviate from its flat counterpart $4Dt$, let us define an effective, time-and geometry-dependent, diffusion coefficient $D^{\mathcal{O}}_{\mathbb{M}}\equiv\left<\mathcal{O}\right>/4t$, where $\mathcal{O}$ is $s^2$, $\delta{\bf R}^2$ or $\delta{\bf R}^{2}_{\perp}$. 
\begin{figure}[h]
\label{fig0}
 \begin{center}
\includegraphics[width=0.23\linewidth]{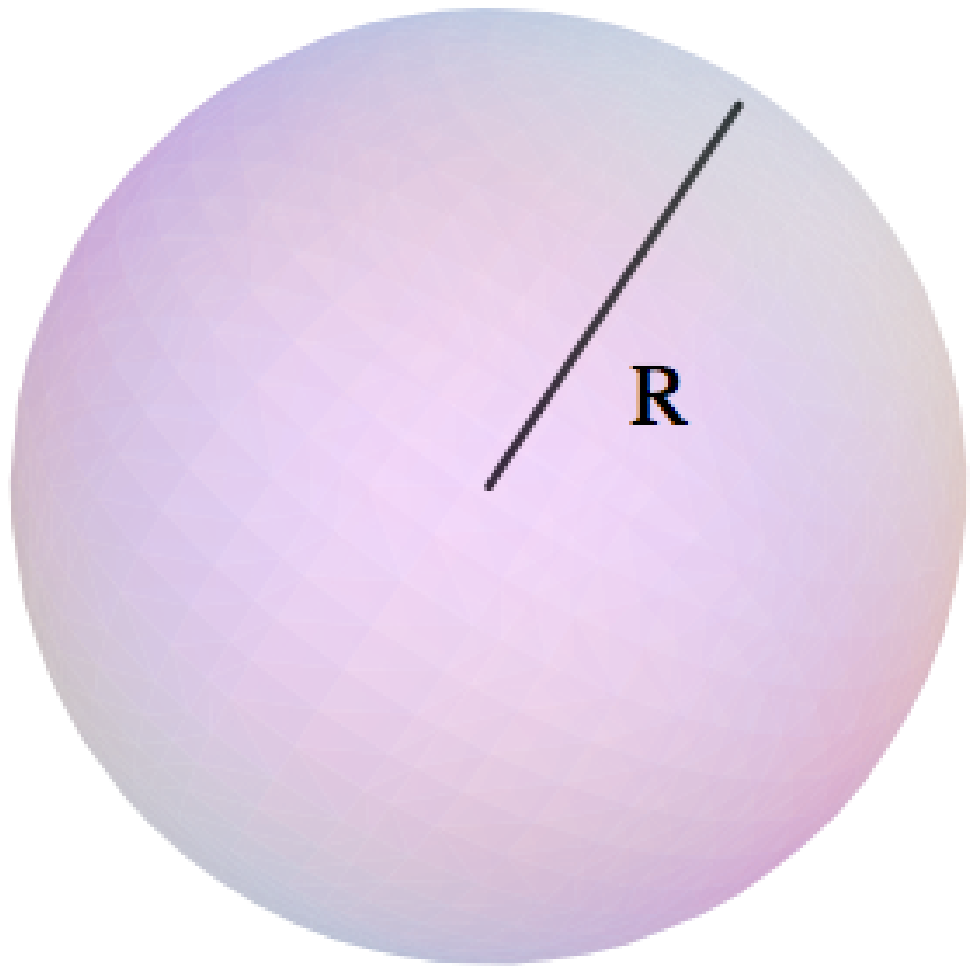}~~~~~~~~~\includegraphics[width=0.37\linewidth]{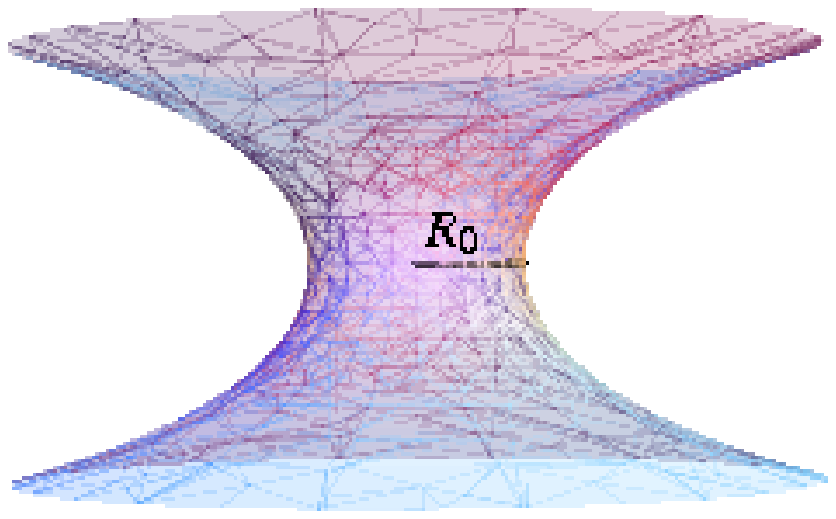}~~~~~~\includegraphics[width=0.32\linewidth]{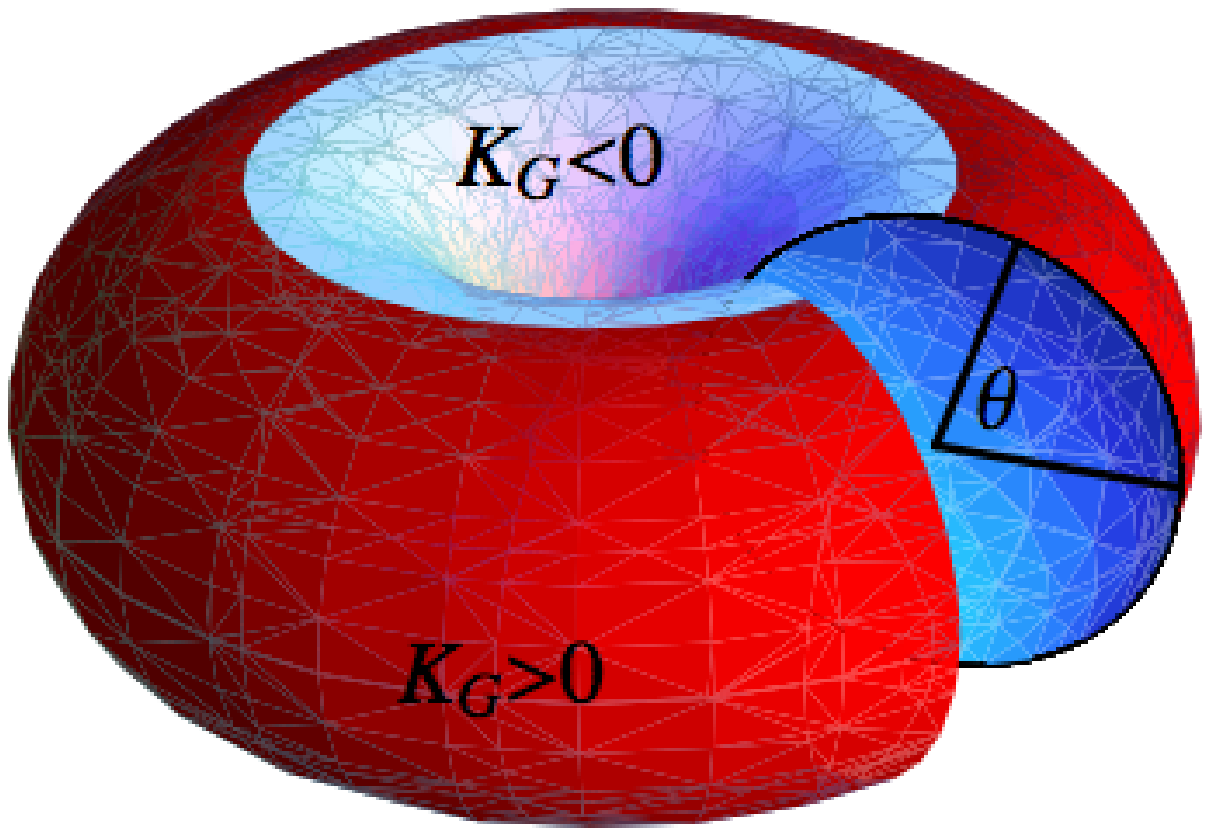}
  \caption{{\small (Color online) A schematic representaction from left to the right of a sphere, catenoid and circular torus is shown.  }}
   \end{center}
 \end{figure}
 
In the case of the sphere $S^2$, the Gaussian and the mean curvature are $K_{G}=1/R^2$ and $K=2/R$, respectively, where $R$ is the radius of the sphere. In this case, from intrinsic, extrinsic and projected points of view the diffusion on the sphere is always below the normal diffusion, that is, $D^{s^{2}}_{S^2}\left(t\right)<D$, $D^{\delta{\bf R}^{2}}_{S^2}\left(t\right)<D$ and $D^{\delta{\bf R}^{2}_{\perp}}_{S^2}\left(t\right)<D$ for all values of time. However, from the intrinsic point of view the Brownian particle travels farther in space than it seems to occur from the extrinsic points of view, in other words, one has  $\left<s^{2}\right>\geq\left<\delta{\bf R}^2\right>\geq\left<\delta{\bf R}^2_{\perp}\right> $ for all values of time, where the equality is satisfied only in the short-time regime. Additionally, in the long-time regime, $\lim_{t\to\infty}$, $\left<s^{2}\right>=\frac{\pi^2-4}{2}R^{2}$ whereas for the extrinsic standpoints one has  $\left<\delta{\bf R}^{2}\right>=2R^{2}$ and $\left<\delta{\bf R}^{2}\right>=\frac{2}{3}R^{2}$.

In the case of minimal surfaces (MS), the Gaussian curvature $K_{g}<0$ and the mean curvature $K=0$. Thus,  $\left<s^{2}\right>$ is non-uniform over the points of the surface. In addition, according to the Eq. (\ref{mean-square}) the diffusion is above the normal diffusion, that is, $D^{s^2}_{{\rm MS}}>D$ at least in the short-time regime. In particular, we study the diffusion on a catenoid, ${\rm Cat}$, which is an example of a minimal surface. In figure (2),  we show the effective diffusion coefficient  $D^{s^2}_{{\rm Cat}}\left(t\right)$  as a function of $Z$ for several values of time. For large values of $Z$, $D^{s^2}_{{\rm Cat}}\left(t\right)\to D$,  having a normal diffusion in agreement with the planar ending of the catenoid.  In contrast, from the extrinsic point of view  $D^{\delta{\bf R}^2}_{{\rm Cat}}=D$ in agreement with {\bf result 1}. 

 \begin{figure}[h]
\begin{center}
\includegraphics[width=0.45\linewidth]{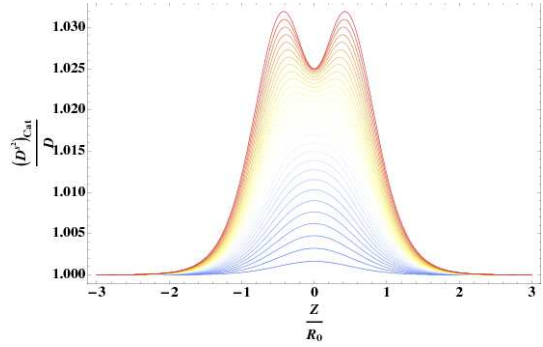}
\label{fig1}
\caption{{\small  (Color online) Free diffusion on a catenoid. The blue hue corresponds to small values of time; the transition to red hue corresponds to an increasing value of time. Dependence of effective diffusion coefficient $D^{s^{2}}_{{\rm Cat}}\left(t\right)$ on $Z$ coordinate is shown.   }}
\end{center}
 \end{figure} 

In the case of circular torus $T^{2}$,  three regions can be defined according to the conditions $K_{G}> 0$, $K_{G} < 0$ and $K_{G}=0$.  In figure (3), we show the effective diffusion coefficient $D^{s^2}_{{\rm T^2}}$ as a function of $\theta$ for several values of time. It is shown that $D^{s^2}_{{\rm T^2}}>D$ in the region where $K_{G}<0$, whereas $D^{s^{2}}_{\rm T^2}<D$ in the region where $K_{G}>0$.  In contrast,  from the extrinsic point of view one has $D^{\delta{\bf R}^{2}}_{{\rm T^2}}<D$ for all regions of $T^2$. 
 \begin{figure}[h]
\begin{center}
\includegraphics[width=0.45\linewidth]{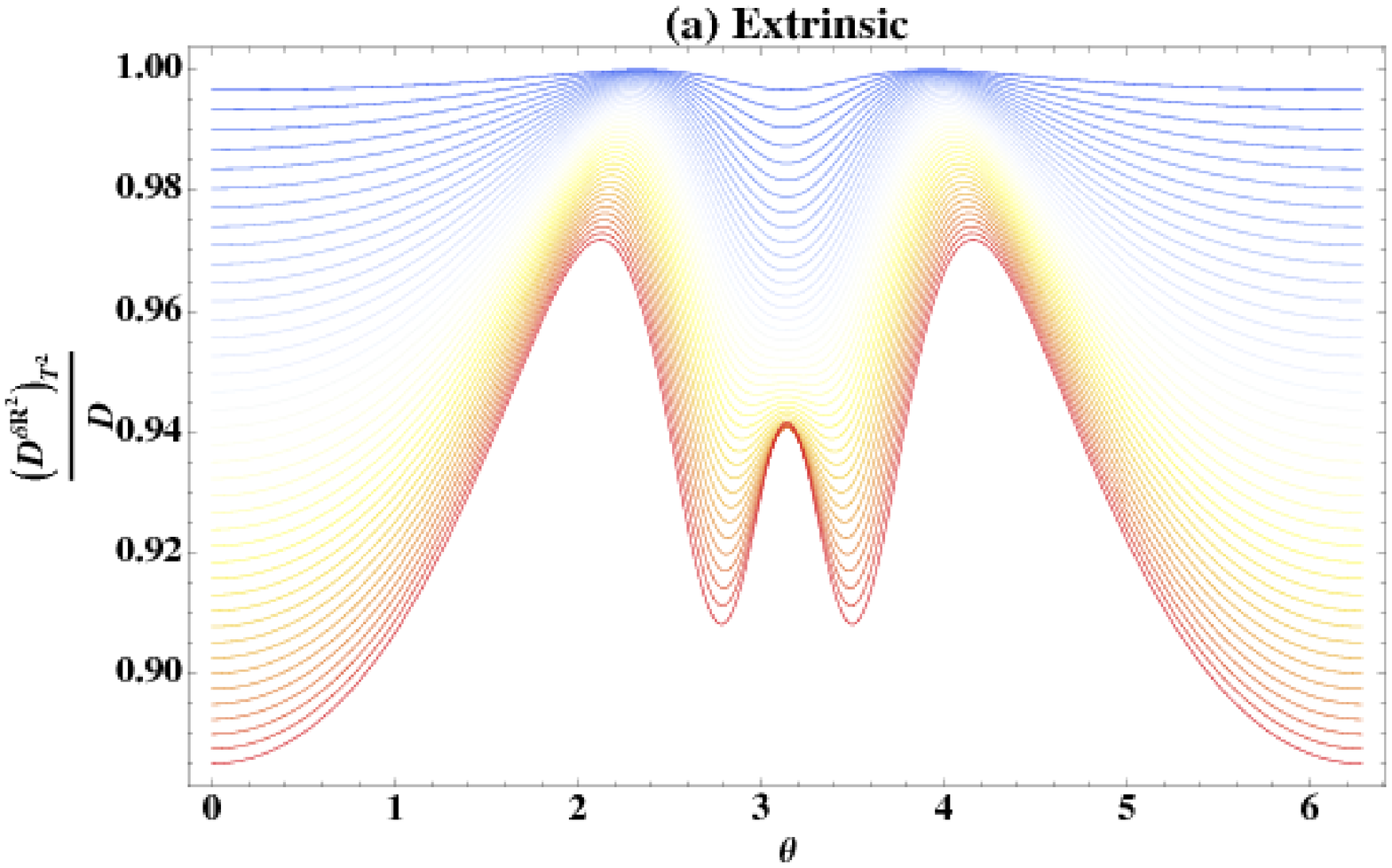}~~~~~\includegraphics[width=0.450\linewidth]{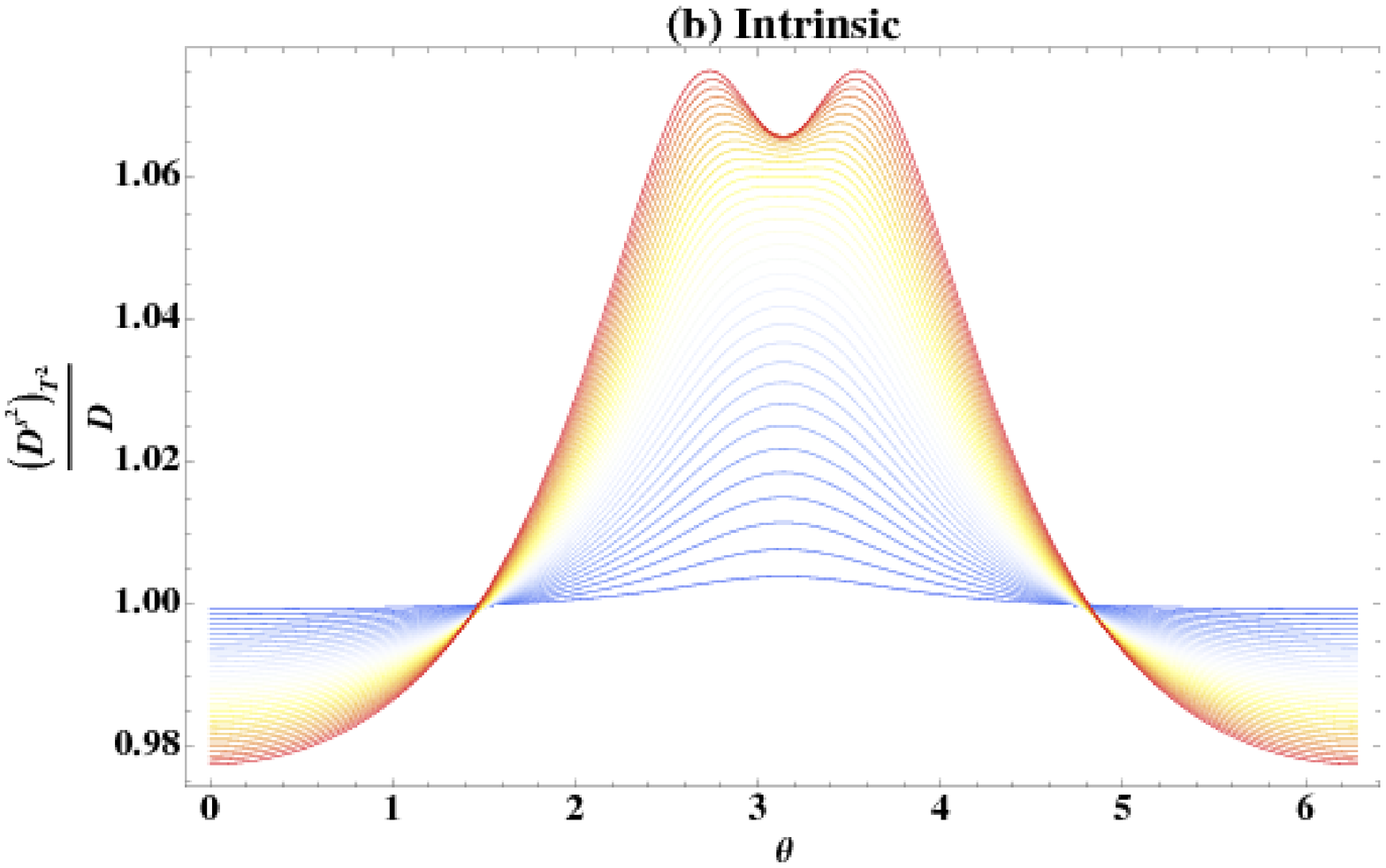}
\label{fig2}
\caption{{\small  (Color online) Free diffusion on a Clifford Torus at short-time. The blue hue corresponds to small values of time; the transition to red hue corresponds to an increasing value of time. (a) Extrinsic point of view of diffusion. Dependence of $D^{\delta{\bf R}^2}_{{\rm T^2}}$  on  $\theta$ coordinate is shown.   (b) Intrinsic point of view of effective diffusion. Dependence  of $D^{s^2}_{{\rm T^2}}$ on $\theta$ coordinate is shown. }}
\end{center}
 \end{figure}

\section{Concluding perspective }

In this paper we  studied Brownian motion over  Euclidean sub-manifolds of dimension $d$. Our approach is based on the Smoluchowski equation on curved manifolds. Here we addressed the question about what functions $\mathcal{O}\left(x\right)$ are useful to describe Brownian motion on curved spaces. In particular, three physical observables are considered for the Brownian displacement, namely, the geodesic displacement $s$, the Euclidean displacement, $\delta{\bf R}$, and the projected displacement, $\delta{\bf R}_{\perp}$. It is noteworthy to mention that the controversy posed in \cite{Faraudo} about the displacement is resolved by considering that all these displacements capture  information of the Brownian motion. 

Here, we study the short-time regime of expectation values of $s^{2}$, $\delta{\bf R} $, $ \delta{\bf R}^2 $, and $ \delta{\bf R}^2_{\perp} $ using the operator method introduced in \cite{Castro-2010}. These observables quantify the curvature effects in different ways. It is shown that Euclidean displacements probe the extrinsic curvature but not the intrinsic one of the embedded surface. In particular, we provide examples to study the diffusion on the sphere, catenoid and Clifford torus. On the one hand, our findings show that from the extrinsic point of view the geometry of the space affects the Brownian motion in such a way that the particle's diffusion is decelerated in contrast with the intrinsic point of view where the dynamics is controlled by the sign of the Gaussian curvature \cite{Castro-2010, Tomoyoshi}. In general, it is remarkable that for sufficiently small time, $ t \ll \tau_ {G} $, the quantities  $ \left <\delta {\bf R}^2\right> $ and $ \left< s^2\right> $ show no curvature effects while $\delta {\bf R}$ does.  Indeed, in the short-time regime the expectation value for $\delta {\bf R}$ is proportional to the scalar curvature $K$ and points outward from the hypersurface with normal direction ${\bf N}$. In other words, it is as if the local curvature would  manifest  through a normal force on the particle. For the mean-square projected displacement, $\left<\delta{\bf R}^2_{\perp}\right>$, formulae are found in terms of the height function of the Monge gauge; in particular, we reproduce the flat diffusion behavior with the projected diffusion coefficient  $D_ {\rm proj}\equiv \frac{D}{d}\left(d-1+N^{2}_{z}\right)$, already obtained, using different methods,  in \cite{Gustaffson}. In addition, we found surprisingly that for the diffusion on minimal surfaces the mean-square Euclidean displacement  exactly matches the usual diffusion corresponding to  flat geometries, although  the Brownian motion has strong intrinsic curvature effects. A very similar effect happens for developable surfaces (with zero Gaussian curvature) at least for short-times \cite{Faraudo}.

Our approach can be extended in various directions. As was mentioned before in \cite{Castro-2010} and by others in \cite{Seifert}, the mobility of proteins or lipids on biomebranes is a complex phenomenon that has several features. In particular, here, we have highlighted the way in which curvature effects appear in the Euclidean displacements. Other features include thermal fluctuations and finiteness of the diffusing particle. In this sense, a natural step forward in this research is to investigate what the role of thermal fluctuations is in the curvature contributions (that are involved in terms of order $\mathcal{O}\left((Dt)^2\right)$ or higher order in the mean-square displacement). In another direction, the simplification occurring in $\left<\delta{\bf R}^2\right>$ and $\left<\delta{\bf R}^2_{\perp}\right>$ for minimal surfaces could be used to probe the structure and topology of lyotropic surfactant phases.

\appendix
\section{Notation}

In this section we review the preliminary notions about  manifolds  (following \cite{Nakahara} and \cite{Spivak}). For sake of generality, let $\mathbb{M}$ be a $d$-dimensional ma\-ni\-fold endowed with a Riemannian metric ${\bf g}:T_{p}\left(\mathbb{M}\right)\times T_{p}\left(\mathbb{M}\right)\to\mathbb{R}$ defined by ${\bf g}=g_{ab}~dx^{a}\otimes dx^{b}$, where  $g_{ab}$ is the metric tensor and $T_{p}\left(\mathbb{M}\right)$ is the tangent space for each $p\in\mathbb{M}$. Also, here, the Riemann tensor is denoted by ${\bf R}=R^{a~~}_{~bcd}~{e_{a}}\otimes dx^{b}\otimes dx^{c}\otimes dx^{d}$\footnote{The set $\left\{e_{a}\right\}$ is a basis for the tangent space and $dx^{a}$ is the corresponding basis in the dual tangent space.}. Using the components of the Riemann curvature tensor we can define   the Ricci tensor $R_{ab}=g^{cd}R_{cadb}$ and scalar curvature $R_{g}=g^{ab}R_{ab}$. In addition, it is convenient to introduce the Laplace-Beltrami operator on scalars  by $\Delta_{g}:C^{2}\left(\mathbb{M}\right)\to\mathbb{R}$ defined by
\begin{eqnarray}
\Delta_{g}~\cdot=\frac{1}{\sqrt{g}}\partial_{a}\left(\sqrt{g}g^{ab}\partial_{b}~\cdot~\right),
\end{eqnarray}
where $g=\det{g_{ab}}$ and $g^{ab}$ is the inverse metric tensor. Also, the derivations are given by $\partial_{a}=\partial/\partial{x}^{a}$, where $x^{a}$, with $a=1, \cdots, d$, are local coordinates of some patch in the manifold. In particular, those $d$-dimensional submanifolds embedded in $\mathbb{R}^{d+1}$ are defined through the embedding functions ${\bf X}:\mathcal{D}\subset\mathbb{R}^{d}\to\mathbb{R}^{d+1}$, which assign  $\left(x_{1},\cdots, x_{d}\right)\to{\bf X}\left(x_{1},\cdots,x_{d}\right)\in \mathbb{R}^{d+1}$. In this case, each vector in $T_{p}\left(\mathbb{M}\right)$ can be spanned by $\left\{{\bf e}_{a}\right\}$, where ${\bf e}_{a}:=\partial_{a}{\bf X}$ are the tangent vectors.  In addition, the 1$^{st}$ Fundamental Form of these manifolds  is ${\rm\bf I}:T_{p}\left(\mathbb{M}\right)\to \mathbb{R}$ defined by ${\rm \bf I}\left(\bf v\right)={\bf v}\cdot {\bf v}$;  thus the metric tensor adopts the simple structure $g_{ab}={\bf e}_{a}\cdot{\bf e}_{b}$. Here,  $\cdot$ is the canonical inner product of $\mathbb{R}^{d+1}$ and $\left|~\cdot~\right|$ is defined as the norm coming from this inner product.

The normal direction to the tangent space is determined by the Gauss map ${\bf N}:\Sigma\to S^{d}$ defined by ${\bf N}^{2}=1$ and ${\bf N}\cdot{\bf e}_{a}=0$ for each $a$. The curvature of this ma\-ni\-fold can be understood in terms of the change of the Gauss map, thus the 2$^{nd}$ Fundamental Form is  ${\rm\bf II}:T_{p}\left(\mathbb{M}\right)\to\mathbb{R}$ defined by ${\rm \bf II}\left({\bf v}\right)=d{\bf N}_{p}\left({\bf v}\right)\cdot{\bf v}$; here the components of this form are the extrinsic curvature tensor $K_{ab}={\bf e}_{a}\cdot\partial_{b}{\bf N}$.  The trace of this tensor is the mean curvature  $K=g^{ab}K_{ab}$. Also, it should be noted that the ``egregium" Gauss theorem implies that the Riemann tensor, $R_{abcd}\equiv K_{ac}K_{bd}-K_{ad}K_{bc}$, depends only on the intrinsic geometry.  The  tangent space and its corresponding normal section change direction for each point $p$ in the manifold. The manner in which this change happens  is captured by the Weingarten-Gauss structure equations
\begin{eqnarray}
\nabla_{a}{\bf e}_{b}=-K_{ab}{\bf N},~~~~~~~~~~~ \nabla_{b}{\bf N}=K_{b}^{~a}{\bf e}_{a},
\label{Weingarten-Gauss}
\end{eqnarray}
where $\nabla_{a}$ is the covariant derivative compatible with the metric $g_{ab}$.
\subsection{Minimal surfaces, Circular Torus, and  Monge gauge}

\subsubsection{Minimal surfaces.} A minimal surface (MS) is a surface that has zero mean-curvature, i.e., $K=0$. Thus, according to the Weingarten-Gauss equations, $\nabla_{a}{\bf e}_{b}=-K_{ab}{\bf N}$, one has in particular that $\Delta_{g}{\bf X}=-K{\bf N}$, then any parametrization of a minimal surface satisfies $\Delta_{g}{\bf X}=0$ \cite{Osserman}. In particular, a catenoid is a minimal surface whose embedding functions are ${\bf X}\left(Z,\varphi\right)=\left(R\left(Z\right)\cos\varphi, R\left(Z\right)\sin\varphi,Z\right)$, where $R\left(Z\right)=R_{0}\cosh Z/R_{0}$, and $\left(Z/R_{0},\varphi\right)\in\left(-\infty,\infty\right)\times\left[0,2\pi\right)$. In particular, one can compute Gaussian curvature using the formulae above \begin{eqnarray}
\label{KGcat}
K_{G}=-\frac{1}{R^{2}_{0}\cosh^{4}\left(Z/R_{0}\right)}.
\end{eqnarray}

\subsubsection{Circular torus.} A circular torus has embedding functions ${\bf X}\left(\theta,\varphi\right)=\left(R\left(\theta\right)\cos\varphi, R\left(\theta\right)\sin\varphi,Z\left(\theta\right)\right)$, where $R\left(\theta\right)=a+r\cos\theta$ and $Z\left(\theta\right)=r\sin\theta$, with $\left(\theta,\varphi\right)\in\left[0,2\pi\right)\times\left[0,2\pi\right)$. This surface is rotationaly invariant, compact and has genus $g=1$. Its Guassian and mean curvature are obtained straightforwardly
\begin{eqnarray}
K_{G}=\frac{\cos\theta}{r\left(a+r\cos\theta\right)}, ~~~~~~~~
K=\frac{1}{r}+\frac{\cos\theta}{\left(a+r\cos\theta\right)}.\nonumber
\end{eqnarray}
 It is worthy to mention that a Clifford torus is obtained when two radii satisfy $a=\sqrt{2}r$. 

\subsubsection{Monge gauge.} A Monge parametrization is given by the embedding functions ${\bf X}({\bf x})=\left({\bf x}, h({\bf x})\right)$, for ${\bf x}\in U\subset\mathbb{R}^2$. Using this parametrization the normal vector is given by ${\bf N}=\left(-\partial h,1\right)/\sqrt{1+\left(\partial h\right)^2}$ and the metric tensor is $g_{ab}=\delta_{ab}-\partial_{a}h\partial_{b}h$. In particular, for Minimal Surfaces $\Delta_{g}{\bf X}=0$, thus in Monge parametrization a piece of MS will satisfy $\Delta_{g}h=0$. 

\section{Green formula and boundary terms}
Let $\phi_{1}$ and $\phi_{2}$ be two differentiable scalar functions on the manifold $\mathbb{M}$, $C^{(2)}\left(\mathbb{M}\right)$, thus the following identity is satisfied
 \begin{eqnarray}
 \label{GreenFormula}
 \int_{\mathbb{M}} dv ~\phi_{1}\Delta_{g}\phi_{2}=\int_{\mathbb{M}} dv~\left(\Delta_{g}\phi_{1}\right)\phi_{2}+\int_{\partial \mathbb{M}} da \left(\phi_{1}\partial_{\nu}\phi_{2}-\left(\partial_{\nu}\phi_{1}\right)\phi_{2}\right), 
 \end{eqnarray}
 where $da$ is the volume element of the boundary, $\partial_{\nu}={\nu}\cdot{\nabla^{a}}$ and $\nu$ is the outer normal vector of the boundary $\partial{\mathbb{M}}$ \cite{Chavel}. This identity is useful to prove the following result concerning the derivations of the mean-values. 

\vskip0.5em
\noindent{\bf 1}. {\it Let $\mathcal{O}:\mathbb{M}\to \mathbb{R}$ be a differentiable function, thus the expectation value of  $\mathcal{O}$, with respect to a probability density $P$,  has the following derivations with respect to time
\begin{eqnarray}
\frac{\partial^{k}\left<\mathcal{O}\left(x\right)\right>}{\partial t^{k}}=D^{k}\int dv~ \Delta^{k}_{g}\mathcal{O}\left(x\right)P\left(x, x^{\prime}, t\right)+ D^{k}\int dv~\nabla_{a}J^{a}_{k},
\label{result0}
\end{eqnarray}
where
\begin{eqnarray}
J^{a}_{k}=\sum^{k}_{j=0}\left\{\left(\Delta^{k-j-1}_{g}\mathcal{O}\right)\Delta^{j}_{g}\nabla^{a}P-\left(\Delta^{k-j-1}_{g}P\right)\nabla^{a}\Delta^{j}_{g}\mathcal{O}\right\}.
\end{eqnarray}}
In order to prove  this result we first  differentiate $\left<\mathcal{O}\left(x\right)\right>$ with respect time, then substitute the diffusion equation (\ref{diff.Eq}). Next, we   use  Green's formula (\ref{GreenFormula}) and substitute the initial condition  (\ref{ini.cond}).  
Note that $J^{a}_{k}$ for each $k$ is a vector field on $\mathbb{M}$ thus, by the divergence theo\-rem, for compact manifolds  the right hand side of Eq. (\ref{result0}) vanishes \cite{Chavel}, except for non-trivial topologies. 
We remark that this expression involves the terms $P$ and $\nabla^{a}P$, then by imposing the mixed Neumann and Dirichlet boundary conditions $\left.P\right|_{\partial\mathbb{M}}=0$ and $\left.\nabla^{a}P\right|_{\partial\mathbb{M}}=0$ we are able to ignore the second term of equation (\ref{result0}).

\section{Geometric identities}\label{ap}

\vskip0.5em
\noindent The following identities are useful for the calculation of the mean-values. This identities can be straightforwardly found them using the Weingarten-Gauss, (\ref{Weingarten-Gauss}), repeatedly.
\begin{eqnarray}
\nonumber\label{id3}
\nabla_{a}\left(K{\bf N}\right)&=&\left(\nabla_{a}K\right){\bf N}+KK_{ab}~{\bf e}^{b}\\
\nonumber\label{id4}
\Delta_{g}{\bf N}&=&\nabla_{a}K^{ab}{~\bf e}_{b}-K^{ab}K_{ab}{~\bf N}\\
\nonumber\label{id5}
\Delta_{g}\left(K{\bf N}\right)&=&\left(\Delta_{g}K-KK_{ab}K^{ab}\right){\bf N}+\left(K\nabla_{a}K^{ab}+2K^{ab}\nabla_{a}K\right){\bf e}_{b}\\
\nonumber
\nabla_{c}\Delta_{g}\left(K{\bf N}\right)&=&\left\{\nabla_{c}\left(\Delta_{g}K-KK_{ab}K^{ab}\right)-\left(K\nabla_{a}K^{ab}+2K^{ab}\nabla_{a}K\right)\right\}{\bf N}\nonumber\\
&+&\left\{\left(\Delta_{g}K-KK_{ab}K^{ab}\right)K_{c}^{~d}\nonumber\right.\\&+&\left.\nabla_{c}\left(K\nabla_{a}K^{ab}+2K^{ab}\nabla_{a}K\right)K^{ad}\right\}{\bf e}_{d}\nonumber\\
\label{id6}
\end{eqnarray}
In particular, identities (\ref{id6}) are useful to determine the mean and mean-square Euclidean displacement, that is, $\delta{\bf R}$ and $\mathcal{O}_{2}\equiv\delta{\bf R}^2$. To wit

\begin{eqnarray}
G^{\mathcal{O}_{2}}_{2}&=&\left.\left[(\Delta_{g}K){\bf N}\cdot\delta{\bf R}-2K\nabla^{a}\left(K_{a}^{~b}{\bf e}_{b}\right)\cdot\delta{\bf R}-4\left(\nabla^{a}K\right)K_{a}^{~b}{\bf e}_{b}\cdot\delta{\bf R}\right.\right.\nonumber\\&-&\left.\left.2K^{2}\right]\right|_{\delta{\bf R}=0}=-2K^{2}\nonumber\\
G^{\mathcal{O}_{2}}_{3}&=&2K^{2}K_{ab}K^{ab}-2K\Delta_{g}K-2\Delta_{g}\left(K^{2}\right)\nonumber\\&-&4\nabla_{b}\left(K\nabla^{a}K_{a}^{~b}+\left(\nabla^{a}K\right)K_{a}^{~b}\right).
\end{eqnarray}

 Now, for the projected displacement, $\delta{\bf R}_{\perp}$ are useful the following identities.
\begin{eqnarray}
\label{id1}
\nabla_{a}\delta{\bf R}_{\perp}={\bf e}_{a}-\nabla_{a}h\hat{\bf k},~~~~~~~~~
\Delta_{g}\delta{\bf R}_{\perp}=-\left(K{\bf N}+\Delta_{g}h~\hat{\bf k}\right)
\end{eqnarray}

\section{Some formulae}

Following \cite{Spivak1},  the remainder, in our case,  of the Taylor expansion acquires the following expression
 \begin{eqnarray}
  \epsilon_{n}\left(t\right)=\frac{\left(Dt\right)^{n+1}}{\left(n+1\right)!}C_{n}\left(x^{\prime},\tau\right),~~~~~~
  C_{n}\equiv\int dv~ \Delta_{g}^{n+1}\mathcal{O}\left(x\right)P\left(x,x^{\prime},\tau\right).
  \label{rem}
 \end{eqnarray}
 Using this expression we define the error of our polynomial approximation to be $\delta_{n}=\epsilon_{n}/C_{n}$. In this sense the time $t<\left(\left(n+1\right)! \delta_{n}\right)^{\frac{1}{n+1}}$, in order to have an error of the order $\delta_{n}$. In addition we are able to prove the following results.

\vskip0.5em\noindent{\bf 2.} {\it Let  $\mathcal{O}\left(x\right)$ be an eigenfunction of Laplace-Beltrami operator  $\Delta_{g}$ with eigenvalue $-\lambda$, then the expectation value of $\mathcal{O}\left(x\right)$ is given by
\begin{eqnarray}
\left<\mathcal{O}\left(x\right)\right>=\mathcal{O}\left(x^{\prime}\right)\exp\left(-\lambda Dt\right)
\label{res1}
\end{eqnarray}}
 {\it Proof.}  It is clear that  $\mathcal{O}\left(x\right)$ fulfills all conditions: $\mathcal{O}\left(x\right)$ is a differentiable function. Indeed, the $k$-th action of $\Delta_{g}$ is given by  $\Delta^{k}_{g}\mathcal{O}\left(x\right)=\left(-\lambda\right)^{k}\mathcal{O}\left(x\right)$. Now, using (\ref{rem}) the remainder $\epsilon_{n}\left(t\right)$ has the following expression  $\epsilon_{n}\left(t\right)=\frac{\left(-\lambda Dt\right)^{n+1}}{\left(n+1\right)!}\left<\mathcal{O}\left(x^{\prime}\right)\right>$ then we have
 \begin{eqnarray}
\left| \epsilon_{n}\left(t\right)\right|\leq M\frac{\left(-\lambda Dt\right)^{n+1}}{\left(n+1\right)!}, 
 \end{eqnarray}
where $M$ is a number independent of $n$. It is elementary that  for  any number $a$ and $\epsilon>0$ we have $a^{n}/n!<\epsilon$  for  sufficiently large value of $n$, therefore $\lim_{n\to\infty}\epsilon_{n}\left(t\right)=0$. Now, using (\ref{formula}) we  get the stated result (\ref{res1}).

\vskip0.5em\noindent{\bf 3.} {\it Let  $\mathcal{O}\left(x\right)\in C^{\left(\infty\right)}\left(\mathbb{M}\right)$ such that $\Delta_{g}\mathcal{O}\left(x\right)=C$ for each point on $\mathbb{M}$, where $C$ is a  real constant, then the expectation value of $\mathcal{O}\left(x\right)$ is given by
\begin{eqnarray}
\left<\mathcal{O}\left(x\right)\right>=\mathcal{O}\left(x^{\prime}\right)+CDt
\label{res2}
\end{eqnarray}}
 \noindent {\it Proof.} It is clear that  $\mathcal{O}\left(x\right)$ fulfill all conditions: $\mathcal{O}\left(x\right)$ is a differentiable function. Indeed, the $n$-th action of $\Delta_{g}$ is given by  $\Delta^{n}_{g}\mathcal{O}\left(x\right)=0$ for $n>1$. In this case the remainder is  $\epsilon_{n}\left(t\right)=0$ for $n>1$. In particular, $\lim_{n\to\infty}\epsilon_{n}\left(t\right)=0$.  Now, using (\ref{formula}) we  get the assertion (\ref{res2}).

\end{document}